\documentclass[12pt,preprint]{aastex}

\usepackage{emulateapj5}
\usepackage{epsf}
\usepackage{graphics}

\newcommand{\tx}[1]{\textrm{#1}}

\newcommand{\kms}{km\,$\,{\rm s}^{-1}$}
\newcommand{\dv}{$R^{1/4}\,$}

\newcommand{\mlu}{M$_\odot$/L$_{B,\odot}$}
\newcommand{\ml}{M$_*$/L$_B$}

\newenvironment{inlinefigure}{
\def\@captype{figure}
\noindent\begin{minipage}{0.999\linewidth}\begin{center}}
{\end{center}\end{minipage}\smallskip}

\newenvironment{inlinetable}{
\def\@captype{table}
\noindent\begin{minipage}{0.999\linewidth}\begin{center}}
{\end{center}\end{minipage}\smallskip}

\medskip

\slugcomment{ApJ, in press}

\shorttitle{The Structure and Dynamics of the Early-Type Lens Galaxy
of 0047$-$281} \shortauthors{Koopmans \& Treu}

\received{2002 May 16}
\begin{document}

\title{The Structure and Dynamics of Luminous and Dark-Matter in the
Early-Type Lens Galaxy of 0047$-$281 at z=0.485$^1$}
\footnotetext[1]{Based on observations collected at W.~M. Keck
Observatory, which is operated jointly by the California Institute of
Technology and the University of California, and with the NASA/ESA
Hubble Space Telescope, obtained at STScI, which is operated by AURA,
under NASA contract NAS5-26555.}

\author{L\'eon V.E.\ Koopmans} 
\affil{California Institute of Technology, 
TAPIR, 130-33, Pasadena, CA 91125}
\author{Tommaso Treu}
\affil{California Institute of Technology, 
Astronomy, 105-24, Pasadena, CA 91125}

\email{leon@tapir.caltech.edu, tt@astro.caltech.edu}

\begin{abstract}

We have measured the kinematic profile of the early-type (E/S0) lens
galaxy in the system 0047$-$281 ($z$=0.485) with the {\sl Echelle
Spectrograph and Imager} (ESI) on the W.M. Keck--II Telescope, as part
of the {\sl Lenses Structure and Dynamics (LSD) Survey}.  The central
velocity dispersion is $\sigma=229\pm 15$~\kms, and the dispersion
profile is nearly flat to beyond one effective radius ($R_e$). No
significant streaming motion is found. Surface photometry of the lens
galaxy is measured from {\sl Hubble Space Telescope} images. From the
offset from the local Fundamental Plane (FP), we measure an evolution
of the effective mass-to-light ratio of $\Delta \log
M/L_B=-0.37\pm0.06$ between $z=0$ and $z=0.485$, consistent with the
observed evolution of field E/S0 galaxies.  (We assume $h_{65}=1$,
$\Omega_{\rm m}$=0.3 and $\Omega_\Lambda$=0.7 throughout.)
Gravitational lens models provide a mass of M$_{\rm
E}=(4.06\pm0.20)\times 10^{11}\,h_{65}^{-1}$~M$_\odot$ inside the
Einstein radius of $R_{\rm E}=(8.70\pm0.07)\,h_{65}^{-1}$\,kpc.  This
allows us to break the degeneracy between velocity anisotropy and
density profile, typical of dynamical models for E/S0 galaxies. We
find that constant M/L model, even with strongly tangential anisotropy
of the stellar velocity ellipsoid, are excluded at $>$99.9\%~CL. The
total mass distribution inside $R_{\rm E}$ can be described by a
single power-law density profile, $\rho_t\propto r^{-\gamma'}$, with
an effective slope $\gamma'=1.90^{+0.05}_{-0.23}$ (68\%~CL; $\pm0.1$
systematic error).  Two-component models yield an upper limit
(68\%~CL) of $\gamma\le 1.55(1.12)$ on the power-law slope of the
dark-matter density profile and a projected dark-matter mass fraction
of $0.41(0.54)^{+0.15}_{-0.05}\left(^{+0.09}_{-0.06}\right)$ (68\%~CL)
inside $R_{\rm E}$, for Osipkov--Merritt models with anisotropy radius
$r_i=\infty(R_e)$. The stellar $M_*/L$ values derived from the FP
agrees well with the maximum allowed value from the isotropic
dynamical models (i.e. the ``maximum-bulge solution'').

The fact that both lens systems 0047$-$281 ($z$=0.485) and MG2016+112
($z$=1.004) are well described inside their Einstein radii by a
constant $M_*/L$ stellar mass distribution embedded in a nearly
logarithmic potential -- with an isotropic or a mildly radially
anisotropic dispersion tensor -- could indicate that E/S0 galaxies
underwent little {\sl structural evolution} at $z \la 1$ and have a
close-to-isothermal total mass distribution in their inner
regions. Whether this conclusion can be generalized, however, requires
the analysis of more systems.  We briefly discuss our results in the
context of E/S0 galaxy formation and cold-dark-matter simulations.
\end{abstract}


\section{Introduction}

To understand the evolution and internal structure of luminous and
dark matter in early-type galaxies (E/S0), we have started the {\sl
Lenses Structure and Dynamics} (LSD) Survey. From an observational
point of view, the LSD Survey aims at measuring -- using the Keck-II
Telescope -- the internal (stellar) kinematics of a relatively large sample of
E/S0 gravitational-lens galaxies in the redshift range $z=0-1$ (see
Treu \& Koopmans 2002, hereafter TK02, for a description of the survey
and its scientific rationale). The stellar kinematic profiles are
combined with constraints from a gravitational-lensing analysis -- in
particular the mass enclosed within the Einstein radius -- in order to
break degeneracies inherent to each method individually. In this way,
we are uniquely able to constrain the luminous and dark mass distribution 
and the velocity ellipsoid of the luminous (i.e. stellar) component.

Several of the main questions about E/S0 galaxies that we aim to
answer with the LSD Survey are: (i) What is the amount of dark matter
within the inner few effective radii ($R_e$)?  (ii) Does the dark
matter density profile agree with the universal profiles inferred from
CDM simulations (Navarro, Frenk \& White 1997, hereafter NFW; Moore et
al.\ 1998; Ghigna et al.\ 2001)? (iii) Is there a universal total
(luminous plus dark-matter) density distribution that well describes
the inner regions of E/S0 galaxies? If so, is it isothermal, as
observed in the local Universe and often assumed in lensing analyses?
(iv) How does the stellar mass-to-light ratio of E/S0 galaxies evolve
with redshift, and does it agree with the evolution of the stellar
populations of field early-type galaxies as inferred from Fundamental
Plane (hereafter FP) measurements (Treu et al.\ 1999, 2001a, 2002,
hereafter T99, T01a, T02; Kochanek et al.\ 2000; van Dokkum et al.\
2001)?  (v) Does the structure of E/S0 galaxies evolve between $z=0$
and 1, or is the evolution of the FP purely an evolution of the
stellar population?  (vi) Is the stellar velocity dispersion tensor
isotropic or anisotropic?

First results from the LSD Survey were presented in two recent papers
(Koopmans \& Treu 2002, hereafter KT02; TK02), where we combined our
measurement of the luminosity-weighted stellar velocity dispersion of
the lens galaxy in MG2016+112 at $z=1.004$ with the mass enclosed
by the Einstein radius as determined from gravitational lens
models (Koopmans et al. 2002). A robust constraint was found on the
slope of the total density profile of the lens galaxy,
i.e. $\gamma'=2.0\pm0.1\pm0.1$ (random/systematic errors) for
$\rho_t\propto r^{-\gamma'}$. In addition, we were able to determine
the stellar mass-to-light ratio and constrain the slope of the
dark-matter halo, leading to a relatively simple self-consistent
picture of the lens galaxy: an old and metal-rich stellar component
embedded in a logarithmic (i.e.~isothermal) potential observed at a
look-back time of $\sim 8$ Gyrs -- remarkably similar to many
present-day E/S0 galaxies. Constant M/L models were ruled out at a
very high confidence level and a significant mass fraction
($\sim$75\%) of dark matter was shown to be present inside the
Einstein radius of about 13.7~kpc.  Unfortunately, given the faintness
of the galaxy at $z$=1, only a luminosity-weighted velocity dispersion
could be obtained and only minimal constraints could be set on the
anisotropy of stellar orbits.

In this paper, we present Keck and HST observations, a
gravitational-lens model, and a dynamical model of the lens galaxy in
0047$-$281 at $z=0.485$ (Warren et al.\ 1996, 1998, 1999). This galaxy
is bright and sufficiently extended that we were able to measure a 
spatially resolved velocity dispersion profile, thus setting
unprecedented constraints on the orbital structure of a galaxy at
a look-back time of $\sim 5$~Gyrs. 

The paper is organized as follows. In Section~2, we describe archival
HST observations and spectroscopic observations with ESI on the
Keck--II telescope. In Section 3, we analyze these observations and
determine the luminosity evolution of the lens galaxy with respect to
the local Fundamental Plane. In Section~4, we present a gravitational
lens model from which we determine the mass enclosed by the Einstein
radius.  In Section~5, we present a model for the luminous and
dark-matter distributions of the lens galaxy. In Section~6, the
results from the dynamical models are presented. Section~7 summarizes
and discusses the results.

In the following, we assume that the Hubble constant, the matter
density, and the cosmological constant are
H$_0=65h_{65}$~km\,s$^{-1}$\,Mpc$^{-1}$, $\Omega_{\rm m}=0.3$, and
$\Omega_{\Lambda}=0.7$, respectively. Throughout this paper, $r$ is
the radial coordinate in 3-D space, while $R$ is the radial coordinate
in 2-D projected space.

\section{Observations}

\subsection{Hubble Space Telescope Imaging}

\setcounter{footnote}{1}
Wide Field and Planetary Camera 2 (WFPC2) images of the system are
available from the HST archive, through filters F555W and 
F814W\footnote{Obtained as part of the CASTLeS Survey}. In particular, the
system has been imaged for 9500s in F555W on the Wide-Field Camera,
and 2700s in F814W on the Planetary Camera.

\begin{inlinefigure}
\begin{center}
\resizebox{\textwidth}{!}{\includegraphics{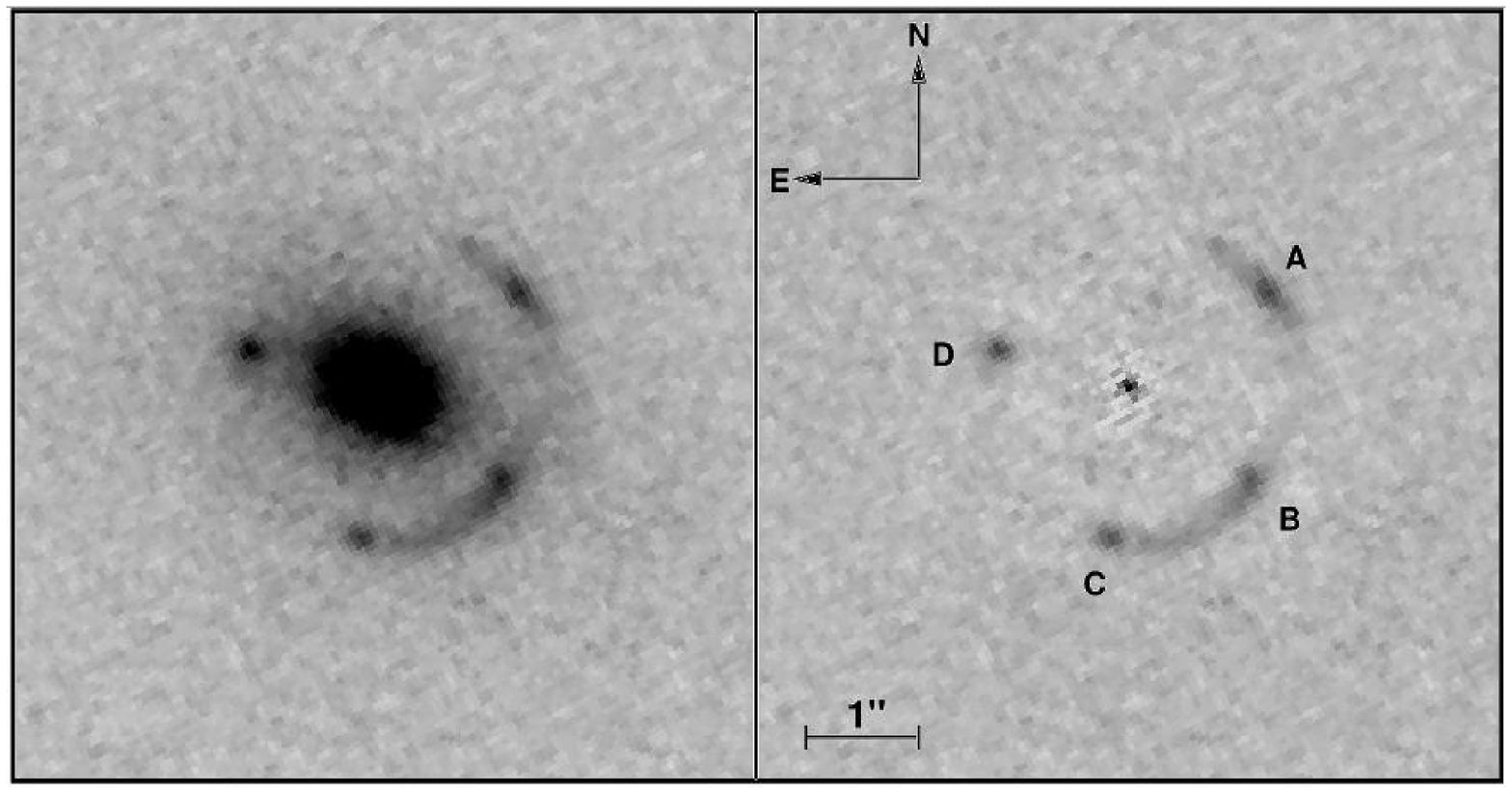}}
\end{center}
\figcaption{\label{fig:HST} Left: HST F555W image of 0047$-$281. 
The images are designated A, B, C and D from the upper right 
going clockwise. Right: Same image with a smooth galaxy model
subtracted. Note the almost continuous ring structure including 
images A, B, and~C.}
\end{inlinefigure}

The images were reduced using a series of {\sc iraf} scripts based on
the {\sc iraf} package {\sc drizzle} (Fruchter \& Hook 2002), to align
the independent pointings and perform cosmic ray rejection. A
subsampled (pixel scale $0\farcs05$) image was produced for the F555W
image. The final ``drizzled'' image in F555W is shown in
Fig.\ref{fig:HST}. Note that for the redshift of the source,
$z=3.595$, Ly-$\alpha$ is redshifted to 5589~\AA\ (Warren et al.\
1998) and the F555W magnitude therefore includes the bright
Ly-$\alpha$ emission.  The F814W image is significantly shallower than
the F555W image and although adequate for measuring the structural
parameters, it is not particularly useful for the lens modeling, since
the multiple images are faint and their positions cannot be accurately
determined.

Surface photometry was performed on the F555W and F814W images as
described in T99 and Treu et al.\ (2001b; hereafter T01b). The galaxy
brightness profiles are well represented by an \dv\ profile, which we
fit -- taking the HST point spread function into account -- to obtain
the effective radius ($R_e$), the effective surface brightness
(SB$_e$), and the total magnitude. The relevant observational
quantities of galaxy G in 0047$-$281 and their errors are listed in
Table~1. Note that errors on SB$_e$ and $R_e$ are tightly correlated
and that the uncertainty on the combination $\log R_e - 0.32 SB_e$
that enters the Fundamental Plane (see Section~\ref{sec:FP}) is very
small ($\sim 0.015$; see Kelson et al. 2000; T01b; Bertin, Ciotti \&
del Principe 2002).  In the right hand panel of Fig.\ref{fig:HST} we
show the HST image after removal of a smooth model for the lens
galaxy. Notice the nearly circular structure of the lensed image
configuration around the lens galaxy.

Rest frame photometric quantities listed in Table~1 -- computed as
described in T01b -- are corrected for galactic extinction using
E(B--V)=0.016 from Schlegel, Finkbeiner \& Davis (1998).

\subsection{Keck Spectroscopy}\label{keck}

We observed 0047$-$281 using the Echelle Spectrograph and Imager (ESI;
Sheinis et al.\ 2002) on the W.M.~Keck--II Telescope during four
consecutive nights (23--26 July, 2001), for a total integration time
of 20,700s (7x1800s+3x2700s). The seeing was good
($0\farcs6<$FWHM$<0\farcs8$) and three out of four nights were
photometric. Between each exposure, we dithered along the slit to
allow for a better removal of sky residuals in the red end of the
spectrum.  The slit (20$''$ in length) was aligned with the major axis
of the galaxy. The slit width of $1\farcs25$ yields an instrumental
resolution of 30\,\kms\ which is adequate for measuring the stellar
velocity dispersion and removing narrow sky emission lines. The
centering of the galaxy in the slit was constantly monitored by means
of the ESI viewing camera (the galaxy was bright enough to be visible
in a few seconds exposure) and we estimate the centering perpendicular
to the slit to be accurate to $\la0\farcs1$.

\medskip
\begin{inlinetable}
\centering
\begin{tabular}{lr}
\hline
\hline
 Redshift (G)          & 0.485$\pm$0.001 \\
 $F814W$ (mag)      & 18.67$\pm$0.09 \\
 $F555W$ (mag)     & 20.61$\pm$0.09 \\
 SB$_{e,F814W}$  (mag/arcsec$^2$)& 20.23$\pm$0.22 \\
 SB$_{e,F555W}$  (mag/arcsec$^2$)& 22.59$\pm$0.07 \\
 $R_{e,F814W}$ (arcsec) & 0.82$\pm$0.12 \\ 
 $R_{e,F555W}$ (arcsec)& 0.99$\pm$0.04 \\
 $b/a$=$(1-e)$ & 0.80$\pm$0.10 \\
 Major axis P.A. ($^\circ$) & $67\pm5$\\
\hline
 $\sigma$  (km\,s$^{-1}$) &	229$\pm$15 \\
 $M_{V}-5\log h_{65}$ (mag)  & $-$22.90$\pm$0.04 \\
 $M_{B}-5\log h_{65}$ (mag)  & $-$22.22$\pm$0.11 \\
 $R_{e,V}$ ($h_{65}^{-1}$kpc) & 5.21$\pm$0.72\\
 $R_{e,B}$ ($h_{65}^{-1}$kpc) & 5.82$\pm$0.58\\
 SB$_{e,V}$  (mag/arcsec$^2$)& 19.29$\pm$0.22 \\
 SB$_{e,B}$ (mag/arcsec$^2$) & 20.39$\pm$0.11 \\
\hline
\hline
\end{tabular}
\end{inlinetable}

\noindent{Table~1.--- \small Observed spectro-photometric quantities
of the lens galaxy (G) in Q0047-281.  The second part of the table
lists rest-frame quantities, derived from the observed quantities as
described in Section~2. Note that $\sigma$ is the central velocity
dispersion corrected to a circular aperture of radius R$_e/8$.
All quantities in this table assume
H$_0=65$~km\,s$^{-1}$\,Mpc$^{-1}$, $\Omega_{\rm m}=0.3$, and
$\Omega_{\Lambda}=0.7$.
\label{tab:HST}}\medskip

Data reduction was performed using the {\sc iraf} package
EASI2D\footnote{developed by D.~Sand and T.~Treu; Sand et al. (2002),
in prep.} as described in KT02. To preserve most of the spatially
resolved information and to achieve an adequate signal-to-noise ratio
at the largest distance for the center of the galaxy, we defined 5
apertures along the spatial direction of the spectrum 
and summed the signal within
each aperture. The apertures correspond approximately to angular
dimensions along the slit: $0\farcs6\times1\farcs25$,
$0\farcs6\times1\farcs25$, $0\farcs4\times1\farcs25$,
$0\farcs6\times1\farcs25$, $0\farcs6\times1\farcs25$ (centered
respectively at $-1\farcs1,-0\farcs5,0'',0\farcs5,1\farcs1$ along the
major axis).  A velocity dispersion profile was measured with the
Gauss--Hermite Pixel-Fitting Software (van der Marel 1994) and the
Gauss Hermite Fourier Fitting Software (van der Marel \& Franx 1993)
on the spectral region including the G-band ($\sim 4304$~\AA), using
as kinematic templates spectra of G--K giants observed at twilight
with a $0\farcs3$ slit width, appropriately smoothed to match the
instrumental resolution of the $1\farcs25$ slit. The two codes provide
consistent measurements. The total error on velocity dispersion was
estimated by adding in quadrature the formal uncertainty given by the
codes, the scatter in the results obtained with different templates
and the semi-difference of the results obtained with the two
codes. The fit to the spectrum from the aperture including part of the
brightest lensed images (A and B) was poor, with severe mismatch and
unstable measurements. We interpret this as due to contamination by
emission from the lensed images. We discarded this measurement from
the analysis. The velocity dispersion profile was then folded around
the center, as determined by fitting the centroid of the light
distribution at the wavelength of the G band, and the velocity
dispersion averaged in the corresponding apertures (symmetric
apertures provided results within the errors). The final results are
listed in Table~2.

\medskip
\begin{inlinetable}
\centering
\begin{tabular}{lcc}
\hline
\hline
Aperture ($\sq''$) & $\sigma$ (\kms) & $\Delta\sigma$ (\kms)\\
\hline
(0.0--0.2)$\times$1.25 & 219 & 12 \\
(0.2--0.8)$\times$1.25 & 212 & 14 \\
(0.8--1.4)$\times$1.25 & 205 & 13 \\
\hline
\hline
\end{tabular}
\end{inlinetable}

\noindent{Table~2.--- \small Kinematic data along the major axis of
0047$-$281. The adjacent rectangular apertures are indicated. The
seeing was 0\farcs7 during the observations.\label{tab:sigma}}\medskip

Using the procedure detailed in T01b, the value in the central
aperture can be corrected to standard central velocity dispersion of
$\sigma=229\pm15$ \kms\, within a circular aperture of radius $R_e/8$
(see also Sec.6.1).  No evidence for significant streaming motions
(e.g. rotation) was found, with an upper limit of 50 \kms\, relative
radial velocity between the center and the outermost aperture.

\section{The Fundamental Plane and the evolution of stellar populations}
\label{sec:FP}

Early-type galaxies in the local Universe occupy approximately a plane
in the three-dimensional space defined by the parameters effective
radius ($\log R_e$), effective surface brightness (SB$_e$) and central
velocity dispersion ($\log \sigma$),
\begin{equation}
\label{eq:FP} 
\log R_{\tx{e}} = \alpha \log~\sigma + \beta~SB_{\tx{e}} + \gamma_{\rm FP} 
\end{equation} 
\noindent
known as the Fundamental Plane (hereafter FP; Dressler et al.\ 1987;
Djorgovski \& Davis 1987).

In recent years, it has been shown that a similar correlation between
those observables exists in clusters out to $z\sim0.8$ (e.g. van
Dokkum \& Franx 1996; Kelson et al.\ 1997; Bender et al.\ 1998; Pahre
1998; van Dokkum et al.\ 1998; J{\o}rgensen et al.\ 1999; Ziegler et
al.\ 2001). The observed evolution of the intercept $\gamma_{\rm FP}$ of the FP
with redshift is consistent with the expectations of passive evolution
models for an old stellar population (redshift of formation $z_f\ga
2$; see e.g. van Dokkum et al.\ 1998). No evidence for a dramatic
evolution of the slopes $\alpha$ and $\beta$ of the FP with redshift
is found with the available data (see J{\o}rgensen et al.\ 1999,
Kelson et al.\ 2000, and T01a for discussion). The correlation is
observed to be tight also in intermediate redshift field samples,
although a faster evolution of the intercept is found in the highest
redshift field samples (to $z\sim0.7$) and interpreted as evidence for
secondary episodes of star formation in the field population at $z<1$
(T02; see also T99, T01a, Kochanek et al.\ 2000, van Dokkum et al.\
2001, Trager et al.\ 2000).

Assuming that galaxy G in 0047$-$281 lies on a FP with slopes similar
to those in the local Universe -- and pure luminosity evolution -- we
can determine the offset of its effective mass-to-light ratio ($M/L
\propto \sigma^2 10^{-0.4SB_e}/R_e$) with respect to the local
relation, which is related to the evolution of the intercept by
$\Delta \log M/L = - 0.4 \Delta \gamma_{\rm FP} / \beta$. As the local FP in
the B band we adopt the relation found by Bender et al.\ (1998),
i.e. $\alpha=1.25$, $\beta=0.32$, and $\gamma_{\rm FP}=-8.895-\log h_{50}$. In
this way, we obtain $\Delta \log M/L_B=-0.37\pm0.06$. The error is
dominated by the observed FP parameters of galaxy G and dominates
uncertainties on the local FP relation.  In Fig.\ref{fig:FP} we plot
the evolution of the effective M/L for cluster and field E/S0 galaxies
as function of redshift (dashed and solid line; from T02), together
with the value obtained for galaxy G (large filled square). The
effective M/L evolution for galaxy G is consistent with what is
observed for field galaxies, i.e. faster than for the cluster sample,
possibly indicating younger luminosity-weighted stellar populations
(see, e.~g., T01a, T02). As described and discussed in TK02, we can
use this measurement to infer the stellar mass-to-light ratio
($M_*/L_B$) of galaxy G assuming that
\begin{equation}
\log (M_*/L_B)_{z}= \log (M_*/L_B)_{0} + \Delta \log (M/L_B),
\label{eq:FPev}
\end{equation}
where the second term on the right hand side of the equation is
measured from the evolution of the FP, and the first term on the right
hand side of the equation can be measured for local E/S0
galaxies. Note that Equation~\ref{eq:FPev} uses the non-trivial
assumption that the stellar mass is a redshift-independent function of
the combination of observables used to define the effective mass
$\sigma^2R_e$ (for a full discussion see T02, and TK02). Using the
local value of $(7.3\pm2.1) h_{65}$~\mlu\ determined from data by
Gerhard et al.\ (2001) as in TK02, we infer $M_*/L_B=(3.1\pm1.0)
h_{65}$~\mlu\ for galaxy G.

\begin{inlinefigure}
\begin{center}
\resizebox{\textwidth}{!}{\includegraphics{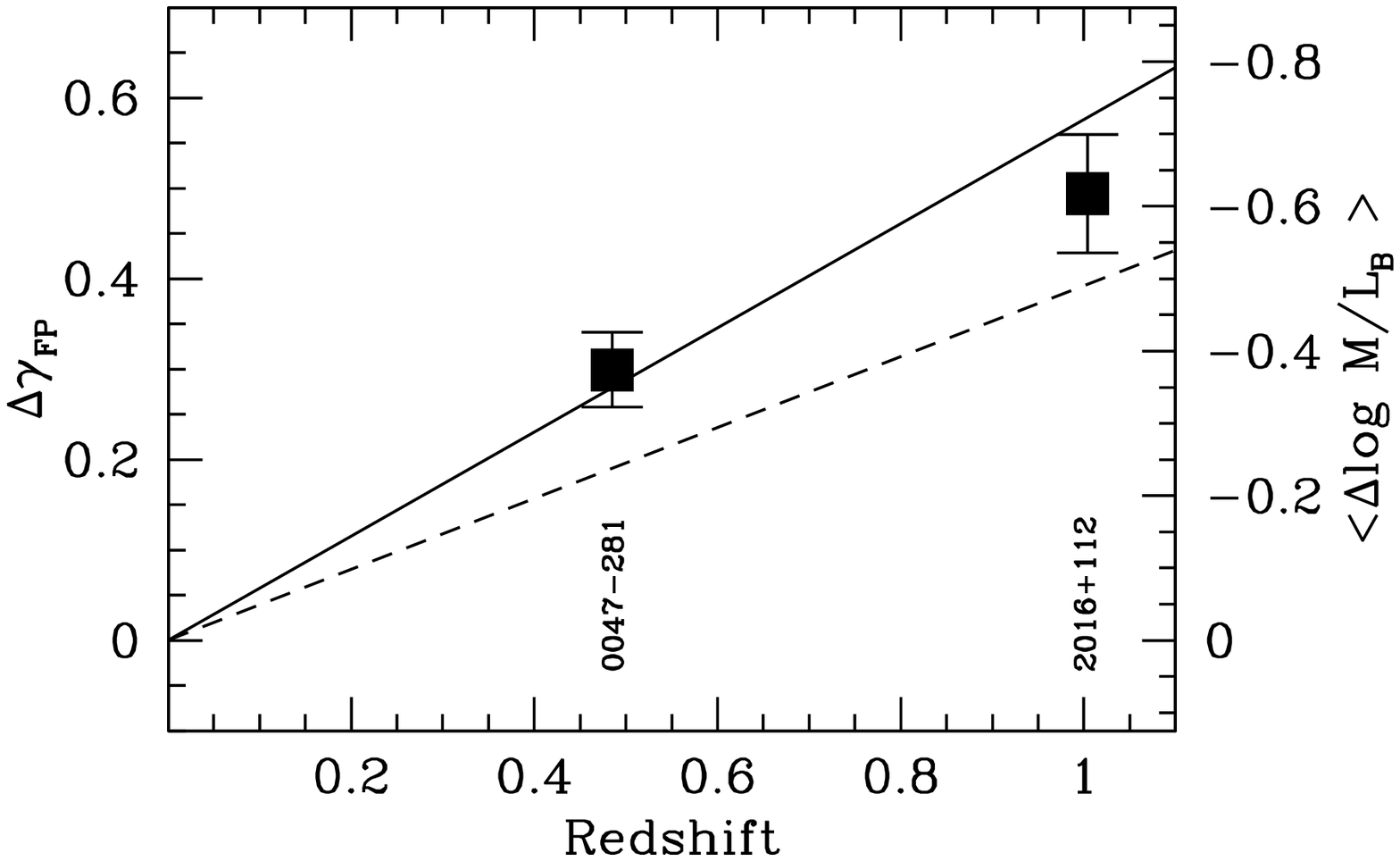}}
\end{center}
\figcaption{Evolution of the M/L as inferred from the evolution of the
FP. The solid and dashed lines indicate linear fits to the M/L
evolution for field (T02) and cluster E/S0 galaxies (van Dokkum et
al.\ 2001) respectively. The large filled square at $z=1.004$
indicates the M/L evolution of galaxy D in MG2016+112 (KT02) while the
large filled square at $z=0.485$ represents galaxy G in
0047$-$281.\label{fig:FP}}
\end{inlinefigure}

\section{Gravitational--Lens Model}\label{sec:lensing}

The Einstein radius ($R_{\rm E}$) and mass enclosed by the Einstein
radius, i.e. $M_{\rm E}\equiv M(<R_{\rm E})$ are quantities required
in the dynamical models that will be discussed in
Sec.\ref{sect:dyn}. Both $R_{\rm E}$ and $M_{\rm E}$ are very
insensitive to the assumed mass profile (see e.g.~Kochanek 1991),
especially for highly symmetric cases such as 0047$-$281 (see
Fig.\ref{fig:HST}).  For consistency -- to ensure uniform modeling
throughout the LSD survey -- we model the lens galaxy as a Singular
Isothermal Ellipsoid (SIE; Kormann et al.\ 1994; see also Chen,
Kochanek \& Hewitt 1995; Kochanek 1995; Grogin \& Narayan 1996;
Koopmans \& Fassnacht 1999; Cohn et al. 2001; Mu{\~n}oz, Kochanek \&
Keeton 2001; Rusin et al. 2002). Although the assumption of a SIE mass
distribution might not be accurate enough for some applications (see,
e.~g., Saha 2000, Rusin \& Tegmark 2001, Wucknitz 2002), we again
stress that this choice of the mass profile does {\sl not}
significantly bias the determination of the quantities used in our
analysis, i.e.  $M_{\rm E}$ and R$_E$. The reason is that the image
deflection angles in four-image systems like 0047$-$281 are not only
nearly the same, but also that the deflection angle is only a function
of the enclosed mass and therefore is little affected by either the
radial mass profile or the ellipticity of the lens (see \S 3 in
Kochanek 1991 for a more detailed discussion).  Constraints on the
mass density profile obtained in Sec.\ref{sect:dyn} are therefore
virtually {\sl independent} of the choice of lens mass model.

\medskip
\begin{inlinetable}
\centering
\begin{tabular}{cccc}
\hline
\hline
 Image & $\Delta x$ ($''$) & $\Delta y$ ($''$) & $\delta(x,y)$ (mas) \\ 
\hline
 G &   $\equiv$0.000  &  $\equiv$0.000 &  --\\
 A &   $+$1.048  &  $+$0.726 &  10,10\\
 B &   $+$0.896  &  $-$0.802 &  10,10\\
 C &   $-$0.126  &  $-$1.165 &  10,10\\
 D &   $-$1.011  &  $+$0.263 &  10,10\\
\hline
\hline
\end{tabular}
\end{inlinetable}

\noindent{Table~3.--- \small Centroid image positions of 0047$-$281
with respect to the centroid of the primary lens galaxy (G).
\label{tab:lensimages}}\medskip

\begin{inlinefigure}
\begin{center}
\resizebox{0.8\textwidth}{!}{\includegraphics{f3.eps}}
\end{center}
\figcaption{A simple gravitational lens model of 0047$-$281. The
cross-haired ellipse indicates the position, position angle and
ellipticity of the lens galaxy. The thick solid contour is the
critical curve, whereas the dashed contours indicate the time-delay
surface. To guide the eye, the lensed images are indicated by two
contours corresponding to circles in the source plane with 15 and
30-mas radius. The subpanel shows a zoom-in of the caustics. The
source position is indicated by a star.
\label{fig:lmodel}}
\end{inlinefigure}

The centroids of the four lensed images are used as constraints on the
lens model (Table~3), assuming errors of 10~mas, i.e. a fifth of a
drizzled pixel size.  We do not use the flux ratios in our analysis
since they are not only difficult to measure accurately for extended
images, but also change as function of position along the arcs and
could be affected by differential dust extinction (e.g. Surpi \&
Blandford 2002).  The position and position angle of the lens-galaxy
mass distribution are set equal to those of its surface brightness
distribution (Table~1) and the presence of an external shear is
allowed for. Hence, there are six free parameters (i.e. the
ellipticity and velocity dispersion of the lens-galaxy mass
distribution, the source position, and the external shear), eight
constraints (i.e. the four image positions) and consequently two
degrees of freedom.  The velocity dispersion of the SIE mass model is
defined such that the mass enclosed by the critical curve is equal to
that inside the Einstein radius of a Singular Isothermal Sphere with
the same velocity dispersion (Kormann et al.\ 1994).

\begin{inlinefigure}
\begin{center}
\resizebox{\textwidth}{!}{\includegraphics{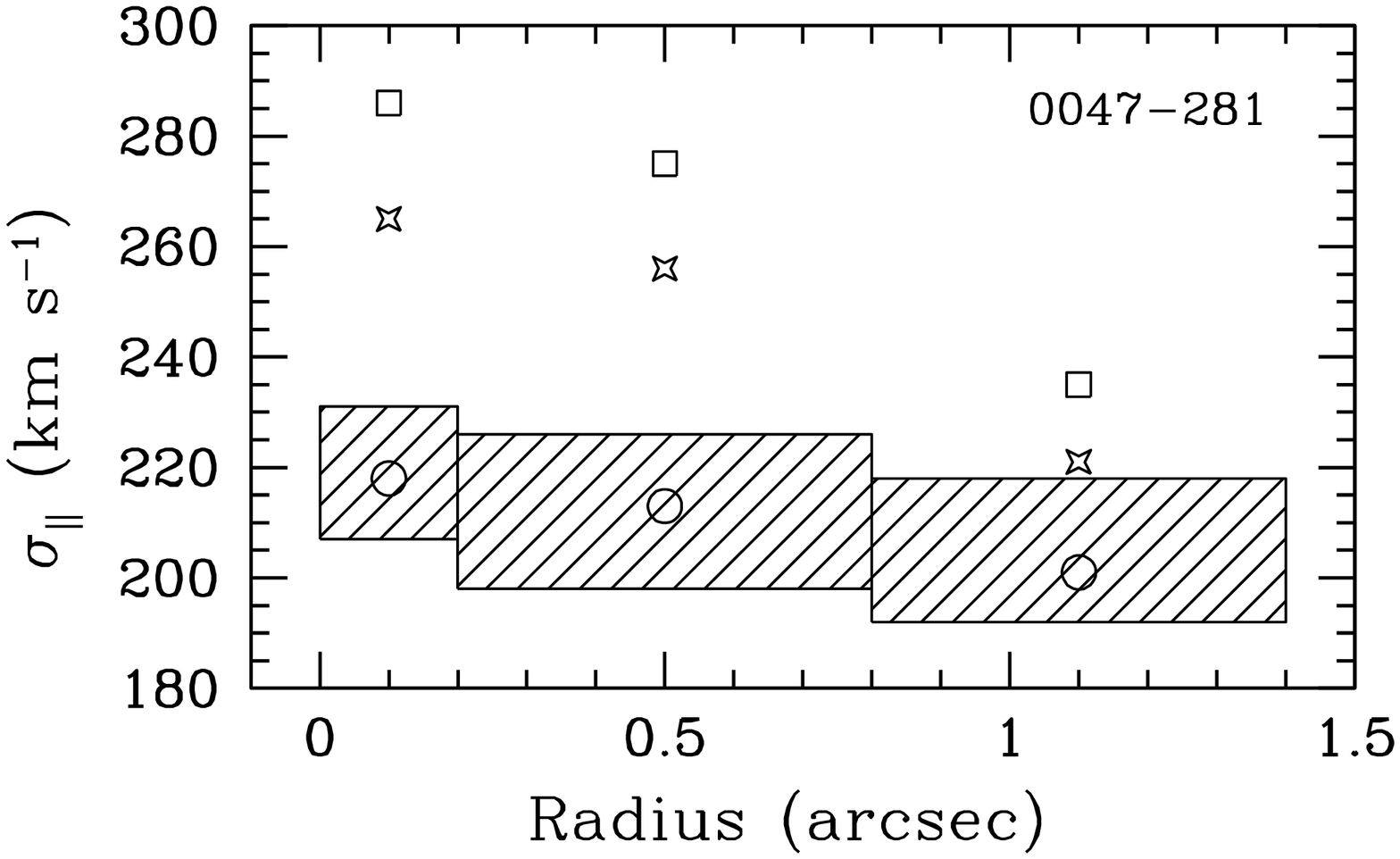}}
\end{center}
\figcaption{Velocity dispersion profile of 0047$-$281 along the major
axis. The box height indicates the 68\% measurement error, whereas the
box width indicates the spectroscopic aperture (times the slit width
$1\farcs25$). The open squares are the corresponding values for an
isotropic constant M/L model, whereas the open stars are for a model
with 10\% lower $M_{\rm E}$ (twice its error) and 10\% higher $R_e$
(both of which lower $\sigma$) with $r_i=\infty$. The
open circles indicate the isotropic Hernquist luminous component
embedded in a total mass distribution with $\gamma'$=1.90.  
See Section~6 for
details. \label{fig:dispersion}}
\end{inlinefigure}

The best solution for a single SIE mass distribution yields
$\sigma_{\rm SIE}=253$~\kms\ for the lens galaxy, whereas the required
external shear is 0.13.  The $\chi^2$ is rather high, 306 for two
degrees of freedom. It is possible to device more elaborate models
(for example with a dwarf companion) that fit the observational
constraints much better. However -- as mentioned before -- the mass
enclosed by the Einstein radius is very insensitive to the precise
lensing model.  As an illustration of the uncertainty related to the
mass modeling we considered including a dwarf companion galaxy with
$\sigma_{SIE}\sim60$ \kms. Although the fit is greatly improved
($\chi^2=0$ because there are more free parameters than constraints),
the velocity dispersion of the primary lens changes by only
$-$4\,\kms.  We will therefore not elaborate further on the details of
the lens model -- which is not the intention of this paper -- and use
the results from SIE mass model in the rest of the analysis. More
detailed lens models are being constructed based on the structure in
the lensed arcs (R.~Webster, private communication).

We note that $\sigma_{\rm SIE}$ is close to the central stellar
velocity dispersion (Table~1). However, we emphasize that $\sigma_{\rm
SIE}$ is a model-dependent expression of the well-determined total
enclosed mass, while the central stellar velocity dispersion depends
on the precise total mass profile, on the luminous mass profile, and
on the dynamical state of the luminous component. Hence, the two
quantities do not have to agree in principle.  Their agreement is
rather an indication of a regular behavior in the physical properties
of early-type galaxies, as discussed in Sections 5, 6 and 7 (see also
Kochanek 1994 and Kochanek et al.\ 2000).

The adopted SIE velocity dispersion corresponds to a circular Einstein
radius of $R_{\rm E}=(8.70\pm 0.07) h_{65}^{-1}$~kpc
(i.e. $\theta_{\rm E}=1\farcs34\pm0\farcs01$) and an enclosed mass of
$M_{\rm E}=4.06\times 10^{11}h_{65}^{-1}$~M$_\odot$. The errors on
$R_{\rm E}$ and $M_{\rm E}$ are correlated (both depend on
$\sigma_{\rm SIE}$): for fixed $R_{\rm E}$ one finds that $\delta
M_{\rm E}/M_{\rm E}= 2\,(\delta\sigma_{\rm SIE}/\sigma_{\rm
SIE})$. This error corresponds to about 3\% for $\delta\sigma=4$~\kms,
i.e. the difference between the model with and without an additional
companion galaxy, which we adopt as systematic error (the random error
on $\sigma_{\rm SIE}$ is a negligible 0.5\%).  By adopting a wide
range of mass profiles and/or ellipticities the enclosed mass changes
by $\la 4\%$ for symmetric four-image systems like 0047$-$281
(e.g. Kochanek 1991). Hence, adding the two contributions in
quadrature, the total error becomes 5\% on $M_{\rm E}$ inside a radius
of $R_{\rm E}\equiv 8.70 h_{65}^{-1}$~kpc.

\section{Dynamical Model}\label{sect:dyn}

Following TK02, we model the galaxy mass distribution as a
superposition of two spherical components, one for the luminous
stellar matter and one for the dark-matter halo\footnote{Spherical
dynamical models provide an adequate approximate description for
computing kinematic quantities of E/S0 galaxies with axis ratio
$b/a\sim0.8$ like 0047-281 (e.~g. Saglia, Bertin \& Stiavelli 1992;
Kronawitter et al.\ 2000), even though clearly they are not
appropriate for the lensing analysis, since, e.~g., they do not
produce quadruply lensed images.}  . The luminous mass distribution is
described by a Hernquist (1990) model
\begin{equation}
\rho_L(r)=\frac{M_* r_*}{2\pi r(r+r_*)^3},
\label{eq:HQ}
\end{equation}
where $M_*$ is the total stellar mass. This profile well reproduces
the \dv surface brightness profile for $r_* = R_e / 1.8153$. In
Sec.\ref{sec:slope}, we also examine the effect of a steeper inner 
core, $\rho_L(r)\propto r^{-2}$, using the 
Jaffe (1983) model. We find the effect to be negligible within the
errors. The dark-matter
distribution is modeled as
\begin{equation}
\rho_d(r)=\frac{\rho_{d,0}}{(r/r_b)^{\gamma}(1+(r/r_b)^2)^{(3-\gamma)/2}}
\label{eq:DM}
\end{equation}
which closely describes a NFW profile for $\gamma=1$, and has the
typical asymptotic behavior at large radii found from numerical
simulations of dark matter halos $\propto r^{-3}$ (e.g. Ghigna et al.\
2000). See TK02 for further discussion of this mass profile and
dynamical model.

According to the CDM simulations given in Bullock et al.\ (2001), a
galaxy with the virial mass of galaxy G at $z=0.485$ has $r_b\approx
50$~kpc~$\gg R_{e} \sim R_{\rm E}$. Hence, in the light of a
comparison with CDM models, we can safely assume that the dark matter
profile in the region of interest (i.e.~inside $\sim R_{\rm E}$) is
well described by a power-law $\rho_{\rm d}\propto
r^{-\gamma}$. Throughout this study, we will set $r_b=50$~kpc
(effectively equal to $r_b=\infty$). The
effects of changing $r_b$ are discussed in TK02.

In addition, we assume an Osipkov-Merritt (Osipkov 1979; Merritt
1985a,b) parametrization of the anisotropy $\beta$ of the luminous
mass distribution
\begin{equation}
\beta(r)=1-\frac{\sigma^2_{\theta}}{\sigma_{r}^2}=\frac{r^2}{r^2+r^2_i},
\label{eq:OM}
\end{equation}
where $\sigma_{\theta}$ and $\sigma_{r}$ are the tangential and radial
component of the velocity dispersion and $r_i$ is called the
anisotropy radius .  Note that $\beta>0$ by definition, not allowing
tangentially anisotropic models. A brief discussion of tangentially
anisotropic models with negative constant values of $\beta$ is given
in Sec.6.2. As a further caveat, note that at infinite radii,
Osipkov-Merrit models become completely radial. Although this
behavior is not commonly found within the inner regions of E/S0
probed by observations (e.~g. Gerhard et al.\ 2001; but see van Albada
1982 and Bertin \& Stiavelli 1993 for theoretical grounds), it has
little impact in the case considered here, since the pressure tensor
only becomes significantly radial well outside the Einstein Radius
and in projection is significantly down-weighted by the rapidly falling 
luminosity--density profile.

The line-of-sight velocity dispersion is obtained solving the three
dimensional spherical Jeans equation (e.g. Binney \& Tremaine 1987 Eq.
4-30) for the luminous component in the total gravitational potential
and computing the luminosity-weighted average along the line of sight
(Binney \& Tremaine 1987; Eq. 4-60; see also, e.~g. Ciotti, Lanzoni
\& Renzini 1996). We correct for the average seeing of 0.7$''$, during
the observations, and average the velocity dispersion -- weighted by
the surface brightness -- inside the appropriate rectangular
apertures. For completeness, we rescale the apertures (Table~2) by 0.9
and 1.1 in the directions of the major and minor axes, respectively,
such that their projection on the axisymmetric model is equivalent to
their projection on an elliptical galaxy with an axial ratio of 0.8,
even though this has minimal effects on the model velocity dispersions
($<1$\%), much smaller than the observational errors. The
uncertainties on seeing, aperture size, and galaxy centering are taken
into account as systematic errors in the following discussion.

\section{Luminous and Dark Matter in 0047$-$281}

The unknown parameters of our dynamical model are $M_*$, $\gamma$,
$r_i$ and $\rho_{d,0}$ (we note again that $r_*=R_e/1.8153$ and
$r_b=50$~kpc). We can eliminate one of these parameters using $M_{\rm
E}$, the mass inside the Einstein radius, which is the most accurately
known constraint on the lens mass model. We choose to eliminate
$\rho_{d,0}$. In addition, we transform $M_*$ into the stellar
mass-to-light ratio $M_*/L_{B}$, fixing the model luminosity exactly
to the value $1.2\times 10^{11}$\,$h_{65}^{-2}$\,L$_{\rm B,
\odot}$. Hence, values of $M_*/L_{B}$ that we derive from the
dynamical model, bears an additional uncertainty of 11\%, i.e. the
observational error on $L_{B}$, whereas $M_*$, which is used in the
model calculation, does not have this error. We use the average
$R_e=(5.52\pm0.55) h_{65}^{-1}$~kpc of the rest-frame V-- and B--band
values.

For any given set $\{M_*/L_{B}, \gamma \}$ the dynamical model is
completely determined and the luminosity-weighted velocity dispersions
for each of the three apertures (Tab.~2) can be computed. The
likelihood is determined assuming Gaussian error distributions and the
confidence contours using the likelihood ratio statistic\footnote{Note
that the likelihood ratio statistic is distributed as a $\chi^2$ only
asymptotically, hence the interpretation of the likelihood ratio
contours as confidence contours is an approximation. However, because
the confidence contours (i.e. Fig.5) on which we base our results only
mildly deviate from true ellipses -- i.e. the limiting case for large
numbers of constraints --, this approximation should work well.}.

\subsection{Power-Law Models}\label{sec:slope}

Before studying the luminous and dark-matter profiles individually, we
determine the effective slope ($\gamma'$) of the total (luminous {\sl
plus} dark) matter density profile ($\rho_t$) inside the Einstein
radius. We emphasize that an effective slope of $\gamma'$ does not
imply that the density profile follows $\rho_t\propto r^{-\gamma'}$
exactly, but only {\sl effectively}.

For MG2016+112 (TK02) we measured an effective slope $\gamma'=2.0\pm
0.1 \pm 0.1$ (random and systematic errors), assuming
$\rho_t\propto r^{-\gamma'}$. Based on the data in Warren et al.\
(1998) and Kochanek et al.\ (2000), we found a similar effective slope
for 0047$-$281. However, the errors were relatively large due to a
lack of an extended kinematic profile and the large uncertainty on the
velocity dispersion given by Warren et al.\ (1998). With the data
presented here we can perform a more accurate measurement, and set
constraints on the anisotropy of the velocity ellipsoid.

In Fig.~\ref{fig:lhplot}(c), we show the likelihood contours of
$\gamma'$ versus the anisotropy radius $r_i$, based on our extended
kinematic profile of 0047$-$281. Three main conclusions can be drawn:
(i) for a spherical isotropic stellar distribution function
($r_i\rightarrow \infty$) $\gamma'=1.90\pm0.05$ (68\% CL) (see also
Fig.\ref{fig:dispersion}); (ii) a lower limit of $r_i/R_e\ga 0.7$
(68\% CL) can be set, implying that the velocity distribution function
is isotropic in the inner regions of the galaxy; (iii) independent of
$r_i$, one finds $\gamma'=1.90^{+0.05}_{-0.23}$ (68\% CL).

To assess systematic errors on $\gamma'$, we varied $M_{\rm E}$ and
$R_{\rm e}$ by the total uncertainties $\pm$5\% and $\pm$10\%,
respectively. The value of $\gamma'$, required to fit the data, 
changes by $\pm$0.05 and
$\pm$0.03, respectively.  Other potential sources of errors, such as
seeing, aperture corrections, aperture offsets, etc., were found to be
negligible. We add both contributions and we conservatively round up
to a total systematic error of~$\pm$0.1, to account for all potential
minor sources of error. 

Although the outer region of galaxy G is well fit by a Hernquist
(1990) model, the inner region with $\rho_L(r)\propto r^{-1}$ is less
constrained due to the finite resolution of the HST images. We
therefore examined power-law models with steeper inner luminosity
density profiles, $\rho_L(r)\propto r^{-2}$, using the model from
Jaffe (1983).  For both $r_i=\infty$ and $r_i=R_e$, we find that the
stellar velocity dispersions change by $\le$\,6~\kms. In general,
Jaffe models give slightly lower velocity dispersions for a fixed
value of $\gamma'$ and the best-fit models therefore yield slightly
higher values of $\gamma'$ (few hundredths) compared with the
Hernquist models. However, the differences are not significant, given
the errors on dispersion profile, and therefore we conclude that our
results are insensitive to the precise shape of the inner luminosity
density profile.

For completeness, we note that our model with $\gamma'=1.90$ results
in a central stellar velocity dispersion of $\sigma$=221~\kms\ inside
$R_e/8$, in excellent agreement with the empirically derived value of
$\sigma$=229$\pm$15~\kms\ in Sec.\ref{keck}, confirming the
self-consistency of our models.

In conclusion, the total mass distribution of galaxy G is well-matched
by a single power-law density profile, that is isothermal
(i.e. $\rho_t\propto r^{-2}$) to within $\sim$5\%, and the velocity
ellipsoid of the luminous component is isotropic at least inside
$\sim70\%$ of the effective radius. Note that this limit is generally
consistent with the limits set by other physical considerations.  In
fact, strongly radial orbits would result in radial instability
(e.g. Merritt \& Aguilar 1985 find that Osipkov-Merritt models with
initial Jaffe density profile are unstable for $r_i\la 0.3r_0$ where
$r_0\approx R_e/0.763$ is the Jaffe half-mass radius; for the effects
of a dark matter halo see, e.g., Stiavelli \& Sparke 1991) or negative
values of the distribution function (e.g. Ciotti 1999, and references
therein).

\begin{inlinefigure}
\begin{center}
\resizebox{\textwidth}{!}{\includegraphics{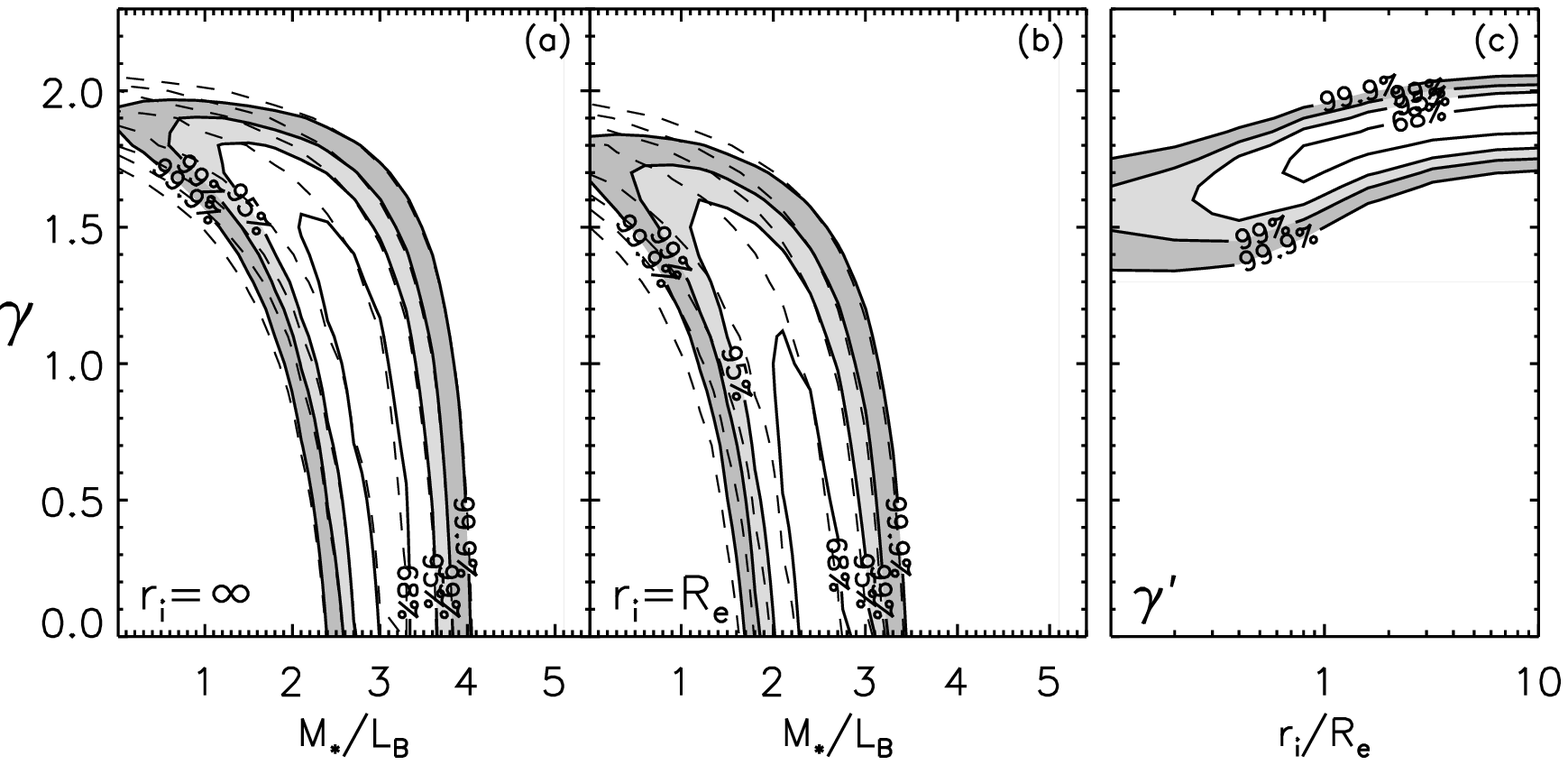}}
\end{center}
\figcaption{Panels {\bf (a)} and {\bf (b)} show the likelihood contours
of the inner slope of the dark-matter halo ($\gamma$) versus the
stellar mass-to-light ratio ($M_*/L_B$) for the lens galaxy in
0047$-$281, for $r_i=\infty$ and $r_i=R_e$, respectively (the mass
enclosed within the Einstein Radius is fixed to M$_E$, see Section~6
for details).  The dashed lines indicate the likelihood contours from
the dynamical model only, whereas the solid lines combine the
constraints from the dynamical model and the FP. Panel {\bf (c)} shows
the likelihood contours of the effective slope ($\gamma'$) versus the
anisotropy radius ($r_i$). All models assume $r_b=50$~kpc. See text
for a more comprehensive description of the models.\label{fig:lhplot}}
\end{inlinefigure}

\subsection{Constant M/L Models}

A stellar mass-to-light ratio of \ml = $(5.4\pm0.8) h_{65}$~\mlu\
is required to account for the mass $M_E$, enclosed by the Einstein
radius.  This is larger than the value \ml = $(3.1\pm1.0)
h_{65}$~\mlu\ derived from the FP evolution. If we had no
kinematic information this could only be interpreted as marginal
evidence for dark matter inside the Einstein radius.

However, the velocity dispersion profile changes the situation
dramatically: no model where mass follows light can be found to fit
the data.  For example, in Fig.\ref{fig:dispersion} we show the
velocity dispersion profile for an isotropic velocity distribution and
constant M/L model (open squares). The dispersion falls too sharply
with radius and this model can be excluded at the $>99.9\%$~CL.  Even
increasing $R_e$ by 10\% and decreasing $M_{\rm E}$ by 10\% (twice its
error) -- both resulting in a smaller stellar velocity dispersion --
the fit (open stars) can still be excluded at the $>$99.9\%
CL. Setting $r_i=R_e$ for the latter model (not shown) worsens the fit and 
the model can be excluded at $>99.9\%$~CL.

Strong tangentially anisotropic models for lens galaxies (see
e.g. Romanowsky \& Kochanek 1999), although probably not very likely,
can lead to flatter velocity dispersion profiles, even if M/L is
constant. We therefore tested models with constant negative 
values of $\beta$. For
$\beta\la -1.5$, we find that the dispersion profile indeed becomes
flat. However, the predicted stellar velocity dispersion is much
higher (i.e. 240--260~\kms) than the observed values and the model is
excluded at $>99.9\%$~CL.  Models with $\beta\la-1.5$ can only fit the
data when $M_{\rm E}$ is lowered by $\sim$30\%. This is at least six
times the error on $M_{\rm E}$ and is incompatible with any acceptable
lens model
of 0047$-$281. We therefore conclude that mass does not follow light
in the lens galaxy G of 0047$-$281, but that M/L must increase with
radius.

This is a key illustration that knowledge of the enclosed mass, $M_E$,
inferred from the gravitational lens models, breaks the
mass--anisotropy degeneracy.

\subsection{Stellar Mass and Dark-Matter Slope}

We now turn to the two-component models described in Section~5, to
assess the density profile of the halo. In Fig.\ref{fig:lhplot} we
show the likelihood contours (dashed lines) as a function of the
stellar mass-to-light ratio (\ml) and inner slope of the dark-matter
halo ($\gamma$).  We choose two representative values for the
anisotropy radius, $r_i=R_e$ and $r_i=\infty$. Note that both the mass
within the Einstein Radius (M$_E$) and $r_i$ are fixed for all models
in Figures~5(a) and 5(b). Hence, for each panel we effectively have
two free parameters and three data points corresponding to the
velocity dispersion profile (a fourth is given by the $M_*/L_B$
measured by the evolution of the FP, Section~\ref{sec:FP}). We find
that: (i) For \ml$\rightarrow 0$, the slope of the dark-matter halo
$\gamma$ approaches $\sim$2, consistent with the findings of
Sec.\ref{sec:slope} (panel c). (ii) For increasing values of \ml, the
total density profile and the velocity dispersion profile become
steeper. Hence, $\gamma$ has to decrease to fit the data.

When we include constraints on \ml\ from the Fundamental Plane
(Sec.3), the limits tighten on both \ml\ and $\gamma$ (solid contours
in Fig.\ref{fig:lhplot}). Remarkably, we find that the stellar
mass-to-light ratio inferred from the FP agrees well with the maximum
stellar mass-to-light ratio allowed by our isotropic dynamical models
(as for MG2016+112; TK02). Such a ``maximum-bulge solution'' is in
that sense equivalent to the ``maximum-disk solution'' for spiral
galaxies (e.g. van Albada \& Sancisi 1986). We note however that the
uncertainties are still considerable.  In addition, the following
limits are found: (i) \ml=$3.2^{+0.2}_{-0.9}\, h_{65}$~\mlu\ and
$\gamma$$<$1.55 (68\%~CL) for $r_i=\infty$ and (ii)
\ml=$2.5^{+0.3}_{-0.5}\,h_{65}$~\mlu\ and $\gamma<1.12$ (68\%~CL) for
$r_i=R_e$. Because the velocity dispersion profile steepens with
increasing radial anisotropy, a smaller $\gamma$ is required to retain
a relatively flat dispersion profile. This explains the mild
degeneracy between $r_i$ and $\gamma$.

In the context of adiabatic contraction (AC) models (e.g. Blumenthal
et al. 1986; Mo, Mao and White 1998), the initial slope ($\gamma_i$)
of the dark-matter halo, i.e. before baryons assembled in the
dark-matter potential well, is in general expected to be shallower
than the observed DM slope ($\gamma$). We find that the difference
between $\gamma_i$ and $\gamma$ is relatively small, because the
stellar mass in the case of 0047$-$280 is quite extended (i.e. large
effective radius) and therefore affects the dark-matter slope less
than found in MG2016+112 (TK02), for example. Even in the absence of
AC, however, we find that $\gamma_i=1.5$ (Moore et al.\ 1998; Ghigna
et al.\ 2001) is inconsistent with the results from 0047$-$281 at the
90\%(68\%) CL for $r_i=R_e(\infty)$, whereas the NFW profile
($\gamma_i=1$; Navarro, Frenk \& White 1997) is consistent at the 68\%
CL for $r_i>R_e$. However, any mechanism, including AC, that steepens
the initial slope by more that
$\Delta\gamma=\gamma-\gamma_i\approx$0.6, would imply that the results
from 0047$-$281 are inconsistent with CDM simulations.

\section{Summary \& Discussion}

We have presented HST and Keck observations of the gravitational lens
system 0047-281. In particular, HST images have been used to measure
the surface photometry of the lens galaxy G and to build a simple lens 
model of the quadruple-image system. Keck-ESI data have been used to measure
a spatially resolved velocity dispersion profile extended beyond the
effective radius, with exquisite accuracy ($\sim 5$\%). We have
combined all these measurements to study the internal structure and
dynamics of the lens galaxy at $z=0.485$, finding the following:

\smallskip\noindent {\bf (i)} The offset of galaxy G from the local
Fundamental Plane, $\Delta \log M/L_B=-0.37\pm0.06$ between $z=0$ and
$z=0.485$, is consistent with what is observed for field E/S0 galaxies
at similar redshift (T02), i.e. somewhat larger than for cluster E/S0
galaxies. In terms of pure luminosity evolution this could be
explained with intermediate age single stellar populations, or -- more
likely -- with secondary episodes of star formation contributing a
fraction of young stars to an old underlying stellar population (see
discussion in T02).

\smallskip\noindent {\bf (ii)} No dark-matter or constant M/L models
are excluded at $>99.9\%$~CL.  Also constant M/L models with strongly
tangential anisotropy of the stellar velocity ellipsoid are excluded
at $>$99.9\%~CL.

\smallskip\noindent {\bf (iii)} The stellar mass-to-light ratio
\ml=$(3.1\pm 1.0)\,h_{65}$~\mlu\ obtained from the offset of the FP is
inconsistent with the required \ml=$5.4\,h_{65}$~\mlu\, to fully
account for $M_E$. This suggest the presence of dark matter. The FP
value is consistent with what is obtained with our two-component
dynamical models, and combining the two constraints we find
\ml=$3.2^{+0.2}_{-0.9}\,h_{65}$~\mlu\ (68\% CL) for an isotropic
velocity ellipsoid.  Hence, dark matter comprises a fraction of
$0.41^{+0.05}_{-0.15}$ of the total mass enclosed by the Einstein
radius of 8.70$\,h_{65}^{-1}\,$kpc for $r_i=\infty$. For $r_i=R_{e}$
this fraction increases to $0.54^{+0.09}_{-0.06}$.  The data are also
consistent with an Osipkov-Merritt (OM) radial anisotropy with
anisotropy radius $r_i\ge 0.7 R_e$ (68 \% CL).

\smallskip\noindent {\bf (iv)} The total (luminous plus dark) mass
distribution inside the Einstein radius can be described by a single
power-law density distribution, $\rho_t\propto r^{-\gamma'}$, with
$\gamma'=1.90\pm0.05$ (68\% CL) for isotropic models,
i.e. $r_i=\infty$. In general, $\gamma'=1.90^{+0.05}_{-0.23}$ (68\%
CL) is found. The systematic error is estimated at 0.1.

\smallskip\noindent {\bf (v)} An upper limit $\gamma\la 1.55$ (68 \%
CL) is found on the slope on the dark-matter halo inside the Einstein
radius for an isotropic model. This limit tightens to $\gamma\la 1.12$
for mildly anisotropic models with $r_i=R_e$. Initial dark-matter
profiles with $\gamma_i=1.5$ (Moore et al.\ 1998; Ghigna et al.\ 2001)
are therefore only marginally acceptable, especially since the profile is
expected to be less steep before the galaxy assembled.  If $\gamma$
steepens by $\Delta\gamma>0.6$ during galaxy formation all CDM
simulations are inconsistent with our results.

\smallskip

In summary, the lens galaxy in 0047$-$281 appears to convey the same picture
formulated for MG2016+112 that early-type galaxies at significant
look-back times can be effectively described by a \dv luminous
component (modeled in this paper as either a Hernquist or Jaffe
profile) embedded in a nearly-isothermal total mass distribution and
that their stellar velocity dispersion is relatively isotropic, in
particular inside the effective radius. In fact, both lens galaxies in
MG2016+112 (KT02, TK02) and 0047$-$281 show that the total mass
distribution is well approximated to within 5\% by a simple power law
density profile $\rho_t\propto r^{-2}$ (i.e. isothermal). 
Whether this conclusion can be generalized, however, requires
the analysis of more systems.

Even more so, we have shown that deviations from isothermality or
isotropy in the lens galaxies of 0047$-$281 and MG2016+112 quickly
lead to inconsistencies with constraints from either the FP, the
observed stellar kinematics, the stellar mass-to-light ratio,
observations of local E/S0 galaxies, the gravitational-lens models,
etc., whereas the models that fit all constraints are internally
consistent, appear to agree with all observational constraints 
available, and indicate both isothermality and near-isotropy.  
Constant M/L or steep mass profiles inside the Einstein radius are 
excluded at very high confidence levels.

A physical explanation is required {\sl if} isotropy and the almost
perfect isothermality are confirmed to be generic features of
early-type galaxies.  In particular, this regularity might suggest
that luminous and dark-matter were strongly coupled at some point
during galaxy assembly. Whereas adiabatic contraction has been
suggested as a mechanism that can lead to near-isothermal mass
profiles (e.g.~Keeton 2001), it is not clear why such a process should
{\sl only} stop when the density profile is isothermal to better than
apparently a few percent (see also TK02). Adiabatic contraction also
leads to a slope of the inner density profile, inconsistent with the
observed absence of lensed images in the centers of lens galaxies
(e.g.~Keeton 2001), if either the central black holes are not very
massive or the inner density profiles do not steepen through some
other process. Violent relaxation could be a natural and viable
explanation for this regularity, although it should also be explained
why luminous and dark matter have {\sl different} density profiles
(see discussion in TK02 and references therein). A combination of the
two processes during some period in the formation of the galaxy can
not be excluded.

The striking similarity of the internal structure of E/S0 galaxies at
large look-back times with the internal structure of local E/S0
galaxies (e.g. Franx et al. 1994; Bertin et al.\ 1994; Rix et
al. 1994; Gerhard et al.\ 2001; see also Kochanek 1995 and reviews by
de Zeeuw \& Franx 1991; Bertin \& Stiavelli 1993; Merritt 1999)
suggests that little structural evolution occurred during the past 8
Gyrs (although again a larger sample is needed to make a quantitative
and general statement). The lack of significant structural evolution
is also suggested by the remarkable agreement between the stellar
\ml\, obtained with our dynamical models and the stellar \ml\,
estimated using the FP evolution.  This fact adds further evidence in
favor of a scenario where the general population of massive (field)
E/S0 galaxies changed little in the past 8 Gyrs (from $z\sim1$) -- as
indicated for example by the modest evolution in their number density
(Schade et al.\ 1999; Im et al.\ 2002; Cohen 2002; McCarthy et al.\
2002) and by the little evolution in the scatter of the FP (T02) --
with most of the evolution being driven by ageing of old stars and
secondary episodes of star formation (Jimenez et al.\ 1999; Trager et
al.\ 2000; Menanteau, Abraham \& Ellis 2001; T02).

{\acknowledgments We thank Eric Agol, Andrew Benson, Giuseppe Bertin,
Roger Blandford, Richard Ellis, Chris Kochanek, and Massimo Stiavelli
for useful comments on this manuscript and stimulating
conversations. We thank the referee for the comments that helped
clarify the presentation of our results. The use of the Gauss-Hermite
Pixel Fitting Software and Gauss-Hermite Fourier Fitting Software
developed by R.~P.~van der Marel and M.~Franx is gratefully
acknowledged. The ESI data were reduced using software developed in
collaboration with D.~Sand.  We acknowledge the use of the HST data
collected by the CASTLES collaboration. LVEK and TT acknowledge
support by grants from NSF and NASA (AST--9900866; STScI--GO
06543.03--95A; STScI-AR-09222). We thank J. Miller, M. Bolte,
R. Guhathakurta, D. Zaritsky and all the people who worked to make ESI
such a nice instrument. Finally, the authors wish to recognize and
acknowledge the very significant cultural role and reverence that the
summit of Mauna Kea has always had within the indigenous Hawaiian
community.  We are most fortunate to have the opportunity to conduct
observations from this mountain.  }

\clearpage

\clearpage

\clearpage

\clearpage

\clearpage

\clearpage

\bigskip

\clearpage 

\bigskip


\begin{thebibliography}{}

\bibitem[Bender et al. 1998]{B98} Bender R., Saglia R.~P., Ziegler B.,
Belloni P., Greggio L., Hopp U., Bruzual G., 1998, ApJ, 493, 529
\bibitem[Bertin, Ciotti \& del Principe]{BCP02} Bertin, G., Ciotti, L., del Principe, M., 2002, A\&A, 386, 149
\bibitem[Bertin \& Stiavelli 1993]{BS93} Bertin, G., \& Stiavelli, M., 1993, Rep. Prog. Phys, 56, 493
\bibitem[Bertin et al. 1994]{B94} Bertin, G. et al.\ 1994, A\&A, 292, 381
\bibitem[Binney \& Tremaine 1987]{BT87} Binney, J. \& Tremaine, S, 1987, Galactic Dynamics, Princeton University Press, Princeton
\bibitem[Blumenthal et al. 1986]{B86} Blumenthal, G.~R., Faber, S.~M., Flores, R., Primack, J.~R. 
\bibitem[Bullock et al.\ 2001]{B01} Bullock, J.~S., Kolatt T.~S., Sigad, Y., Somerville, R.~S., Kravtsov, A.~V., Klypin, A.~A., Primack, J.~R., \& Dekel, A., 2001, MNRAS, 321, 598
\bibitem[Chen, Kochanek, \& Hewitt(1995)]{1995ApJ...447...62C} Chen,
G.~H., Kochanek, C.~S., \& Hewitt, J.~N.\ 1995, \apj, 447, 62
\bibitem[Ciotti (1999)]{C99} Ciotti, L., 1999, \apj, 520, 574
\bibitem[Coehn 2002]{Co02} Cohen, J.~G.  2002, \apj, 567, 672
\bibitem[Cohn, Kochanek, McLeod, \& Keeton(2001)]{2001ApJ...554.1216C}
Cohn, J.~D., Kochanek, C.~S., McLeod, B.~A., \& Keeton, C.~R.\ 2001, \apj,
554, 1216
\bibitem[de Zeeuw \& Franx 1991]{dZF91} de Zeeuw, T., \& Franx, M., 1991, ARA\&A, 29, 239
\bibitem[Djorgovski \& Davis 1987]{DD87} Djorgovski S.~G., Davis M., 1987, ApJ, 313, 59 
\bibitem[Dressler et al.\ 1987]{D87} Dressler, A., Lynden-Bell, D., Burstein, D., Davies, R.~L., Faber, S.~M., Terlevich, R, Wegner G. 1987, ApJ, 313, 42
\bibitem[Falco, Kochanek, \& Munoz(1998)]{1998ApJ...494...47F} Falco, 
E.~E., Kochanek, C.~S., \& Munoz, J.~A.\ 1998, \apj, 494, 47 
\bibitem[Franx et al. 1994]{F94} Franx, M., van Gorkom J.~H., \& de Zeeuw, P.T. 1994, \apj, 436, 642
\bibitem[Fruchter \& Hook 2002]{FH02} Fruchter, A.~S. \& Hook R.~N., 2002, PASP, 114, 144
\bibitem[Gerhard et al. 2001]{G01} Gerhard, O., Kronawitter, A., Saglia, R.~P., \& Bender, R., 2001, \aj, 121, 1936
\bibitem[Ghigna et al.\ 2000]{G00} Ghigna, S., Moore, B., Governato,
F., Lake, G., Quinn, T., Stadel, J., 2000, \apj, 544, 616
\bibitem[Grogin \& Narayan(1996)]{1996ApJ...464...92G} Grogin, N.~A.~\&
Narayan, R.\ 1996, \apj, 464, 92
\bibitem[Helbig et al.(1999)]{1999A&AS..136..297H} Helbig, P., Marlow, D., 
Quast, R., Wilkinson, P.~N., Browne, I.~W.~A., \& Koopmans, L.~V.~E.\ 1999, 
\aaps, 136, 297. 
\bibitem[Hernquist 1990]{H90} Hernquist, L., 1990, \apj, 356, 359
\bibitem[Im et al.\ 2002]{Im02} Im, M., Faber, S.~M., Koo, D.~C.,
Phillips, A.~C., Schiavon, R.~P., Simard, L. \& Willmer, C.~N.~A.,
2002, ApJ, 571, 136
\bibitem[Jaffe(1983)]{1983MNRAS.202..995J} Jaffe, W.\ 1983, \mnras, 202,
995
\bibitem[Jimenez et al.\ 1999]{Ji99} Jimenez, R., Friaca, A., Dunlop, J.~S., Terlevich, R.J., Peacock J.~A., Nolan, L.~A., 1999, MNRAS, 305, L16
\bibitem{JFH99} J{\o}rgensen I., Franx M., Hjorth J., van Dokkum
P.~G., 1999, MNRAS, 308, 833
\bibitem[Keeton (2001)]{K01} Keeton, C.~R. 2001, ApJ, 561, 46
\bibitem[Kelson et al.\ 1997]{KDFIF} Kelson D.~D., van Dokkum P.~G.,
Franx M., Illingworth G.~D., Fabricant D., 1997, ApJ, 478, L13
\bibitem[Kelson et al.\ 2000a]{K2000a} Kelson D.~D., Illingworth
G.~D., van Dokkum P.~G., Franx M., 2000a, ApJ, 531, 137
\bibitem[Kochanek (1991)]{K91} Kochanek, C.~S., 1991, \apj, 371, 289
\bibitem[Kochanek (1994)]{K94} Kochanek, C.~S., 1994, \apj, 436, 56
\bibitem[Kochanek(1995)]{1995ApJ...445..559K} Kochanek, C.~S., 1995,
\apj, 445, 559
\bibitem[Kochanek et al. (2000)]{Koch} Kochanek, C.~S. et al. 2000, ApJ, 543, 131
\bibitem[Koopmans 2001]{Ko01} Koopmans, L.~V.~E., 2001, PASA, 18, 179
\bibitem[Koopmans \& Fassnacht (1999)]{KF99} Koopmans, L.~V.~E. \& Fassnacht, C.~D., 1999, \apj, 527, 513
\bibitem[Koopmans \& Treu (2002)]{KT02} Koopmans, L.~V.~E. \& Treu, T., 2002, \apj, 568, L5 (KT02)
\bibitem[Koopmans et al.\ (2002)]{K02} Koopmans, L.~V.~E., Garrett, M.~A., Blandford, R.~D., Lawrence, C.~R., Patnaik, A.~R., Porcas, R.~W., 2001, \mnras, 334, 39
\bibitem[Kormann et al.\ 1994]{Ko94} Kormann, R., Schneider, P., Bartelmann, M., 1994,  A\&A, 284, 285
\bibitem[Kronawitter 2000]{K00} Kronawitter, A., Saglia, R.~P., Gerhard, O., \& Bender, R. 2000, A\&AS, 144, 53
\bibitem[McCarthy et al. 2002]{Mc02} McCarthy, P.~J. et al.\ 2002, ApJ, 560, L131
\bibitem[Menanteau, Abraham \& Ellis (2001)]{M01} Menanteau, F.,
Abraham, R.~G., Ellis, R.~S. 2001, MNRAS, 322, 1
\bibitem[Merritt 1985]{M85a} Merritt, D. 1985a, \aj, 90, 1027
\bibitem[Merritt 1985]{M85b} Merritt, D. 1985b, \mnras, 214, 25
\bibitem[Merritt 1999]{M99} Merritt, D., 1999, PASP, 111, 129 
\bibitem[Merritt \& Aguilar 1985]{MA85} Merritt, D. \& Aguilar, L.~A. 1985, \mnras, 217, 787
\bibitem[Moore et al.\ 1998]{M98} Moore, B., Governato, F., Quinn, T., Stadel, J. \& Lake, G., 1998, \apj, 499, L5
\bibitem[Mu{\~n}oz, Kochanek, \& Keeton(2001)]{2001ApJ...558..657M} Mu{\~n}oz, 
J.~A., Kochanek, C.~S., \& Keeton, C.~R.\ 2001, \apj, 558, 657 
\bibitem[Navarro et al.\ 1997]{NFW} Navarro, J, Frenk, C.~S., \& White S.~D.~M, 1997, \apj, 490, 493 [NFW]
\bibitem[Osipkov 1979]{O79} Osipkov L.~.P., 1979, Pis'ma Astron. Zh., 5, 77
\bibitem[Pahre 1998]{Pa98} Pahre, M. 1998, PhD Thesis, California
Institute of Technology
\bibitem[Romanowsky \& Kochanek 1999]{RK99} Romanowsky, A.~J. \& Kochanek, C.~S., 1999, \apj, 516, 18 
\bibitem[Rix et al.\ 1997]{R97} Rix, H.~W., de Zeeuw, P.~T, Cretton,
N., van der Marel, R.~P., \& Carollo, C.~M.  1997, \apj, 488, 702
\bibitem[Rusin et al.(2002)]{2002MNRAS.330..205R} Rusin, D., Norbury, M.,
Biggs, A.~D., Marlow, D.~R., Jackson, N.~J., Browne, I.~W.~A., Wilkinson,
P.~N., \& Myers, S.~T.\ 2002, \mnras, 330, 205
\bibitem[Saha(2000)]{2000AJ....120.1654S} Saha, P.\ 2000, \aj, 120, 1654
\bibitem[Saglia et al. 1992]{SBS} Saglia, R.~P., Bertin, G. \& Stiavelli, M. 1992, \apj, 384, 433
\bibitem[Schade et al.\ 1999]{CFRS-Es} Schade D. et al., 1999, ApJ, 525, 31
\bibitem[Sheinis et al. 2002]{Sh02} Sheinis, A.~I., Bolte, M., Epps, H.~W.,  Kibrick, R.~I., Miller, J.~S., Radovan, M.~V., Bigelow, B.C. \& Sutin, B.M. 2002, PASP, 114, 851
\bibitem[Schlegel et al.\ 1998]{EXMAPS} Schlegel, D.~J., Finkbeiner
D.~P., Davis M., 1998, ApJ, 500, 525
\bibitem[Stiavelli \& Sparke 1991]{SS91} Stiavelli, M., \& Sparke, L., 1991, \apj, 382, 466
\bibitem[Surpi \& Blandford 2002]{SB02}Surpi, G., \& Blandford, R.~D., 2002, \mnras, submitted, astro-ph/0111160
\bibitem[Trager et al.\ 2000]{TraI} Trager S.~C., Faber S.~M., Worthey G., Gonzalez J.~J., 2000a, AJ, 119, 1645
\bibitem[Treu \& Koopmans (2002)]{TK02} Treu, T. \& Koopmans, L.~V.~E. 2002, \apj, 575, 87 [TK02]
\bibitem[Treu et al.\ (2001a)]{T01a} Treu, T., Stiavelli, M., Bertin G., Casertano, C., \& M{\o}ller, P. 2001a, \mnras, 326, 237 (T01a)
\bibitem[Treu et al.\ (1999)]{T99} Treu, T., Stiavelli, M., Casertano, C., M{\o}ller, P., \& Bertin G. 1999, \mnras, 308, 1037 (T99)
\bibitem[Treu et al.\ (2002)]{T02} Treu, T., Stiavelli, M., Casertano, C., M{\o}ller, P., \& Bertin, G. 2002, \apj, 564, L13 (T02)
\bibitem[Treu et al.(2001b)]{T01b} Treu, T., Stiavelli, M., M{\o}ller,
P., Casertano, C., \& Bertin G. 2001b, \mnras, 326, 221 (T01b)
\bibitem[van Albada \& Sancisi(1986)]{1986RSPTA.320..447V} van Albada, 
T.~S.~\& Sancisi, R.\ 1986, Royal Society of London Philosophical 
Transactions Series A, 320, 447
\bibitem[van der Marel 1994]{vdm94} van der Marel, R.~P. 1994, MNRAS, 270, 271
\bibitem[van der Marel \& Franx 1993]{gh2} van der Marel R.~P., Franx
M., 1993, ApJ, 407, 525
\bibitem[van Dokkum et al.\ (1998)]{pgd98} van Dokkum, P.~G., Franx, M., Kelson D.~D. \& Illingworth G.~D., 1998, ApJ, 504, L17
\bibitem[van Dokkum et al.\ (2001)]{pgd01} van Dokkum, P.~G., Franx, M., Kelson D.~D. \& Illingworth G.~D., 2001, ApJ, 553, L39
\bibitem[Warren et al. (1996)]{W96} Warren, S.~J., Hewett, P.~C., Lewis, G.~F., M{\o}ller, P., Iovino, A., Shaver P.~A., 1996, MNRAS, 278, 139 
\bibitem[Warren et al. (1998)]{W98} Warren, S.~J., Iovino, A., Hewett, P.~C., Shaver P.~A., 1998, MNRAS, 299, 1215 
\bibitem[Warren et al. (1999)]{W99} Warren, S.~J., Lewis, G.~F., Hewett, P.~C., M{\o}ller, P., Shaver P.~A., \& Iovino, A.\ 1999, A\&A, 343, L35 
\bibitem[Ziegler et al. 2001]{Z01} Ziegler, B.~L., Bower, R.~G.,
Smail, I.~R., Davies, R.~L. \& Lee, D. 2001, \mnras, 325, 1571


\end{thebibliography}
\end{document}